\documentclass[11pt]{article}

\usepackage{graphicx}
\usepackage{natbib}

\def\etal{{et al.\/}\ }
\def\Mpc{$h^{-1}$~{\rm  Mpc}}

\title{Formation of the Supercluster-Void Network}

\author{Jaan Einasto
\\
Tartu Observatory, 61602 T\~oravere, Estonia}

\begin{document}

\maketitle

\begin{abstract}

  A review of the study of superclusters based on the 2dFGRS and SDSS is
  given. Real superclusters are compared with models using simulated galaxies
  of the Millennium Run.  We show that the fraction of very luminous
  superclusters in real samples is about five times greater than in simulated
  samples. Superclusters are generated by large-scale density perturbations
  which evolve very slowly. The absence of very luminous superclusters in
  simulations can be explained either by non-proper treatment of large-scale
  perturbations, or by some yet unknown processes in the very early Universe.

\end{abstract}

\section{Introduction}

Superclusters are the most extensive density enhancements in the
Universe of common origin.  Investigation of large systems of
galaxies was pioneered by the study of the {\em Local Supercluster}
by de Vaucouleurs \cite{deV53}, and by Abell \cite{abell61} using
rich clusters of galaxies by Abell \cite{abell} and Abell et al.
\cite{aco}. Superclusters consist of galaxy systems of different
richness: single galaxies, galaxy groups and clusters, aligned in
chains (J\~oeveer, Einasto \& Tago \cite{jet78}, Gregory \& Thompson
\cite{gt78}, Zeldovich, Einasto \& Shandarin \cite{zes82}).

New deep galaxy surveys, such as the Las Campanas Galaxy Redshift
Survey, the 2 degree Field Galaxy Redshift Survey (2dFGRS, Colless
et al.  \cite{col01}, \cite{col03}) and the Sloan Digital Sky Survey
(SDSS, Adelman-McCarthy et al. \cite{dr4}) cover large areas in the
sky and are almost complete up to fairly faint apparent magnitudes.
Thus these surveys are convenient to detect superclusters using both
galaxy and cluster data.  This possibility has been used by
Basilakos \cite {bas03}, Basilakos et al. \cite{bpr01}, Erdogdu et
al.  \cite{erd04}, Porter and Raychaudhury \cite{pr05}, and Einasto
et al.  \cite{e03a}, \cite{e03b}, \cite{ein05}, \cite{e06a}.

The goal of the present review is to analyse properties of superclusters based
on the 2dF Galaxy Redshift Survey and the Sloan Digital Sky Survey Data
Release 4 by Einasto et al. \cite{e06a}, \cite{e06b} and \cite{e06c}, and to
compare properties of real superclusters with theoretical models. For
comparison we use the superclusters found for the Millennium Run mock galaxy
catalogue by Croton et al. \cite{croton06}, that itself based on the
Millennium Simulation of the evolution of the Universe by Springel et al.
\cite{springel05}.

\begin{figure*}[ht]
\centering
\resizebox{0.49\textwidth}{!}{\includegraphics*{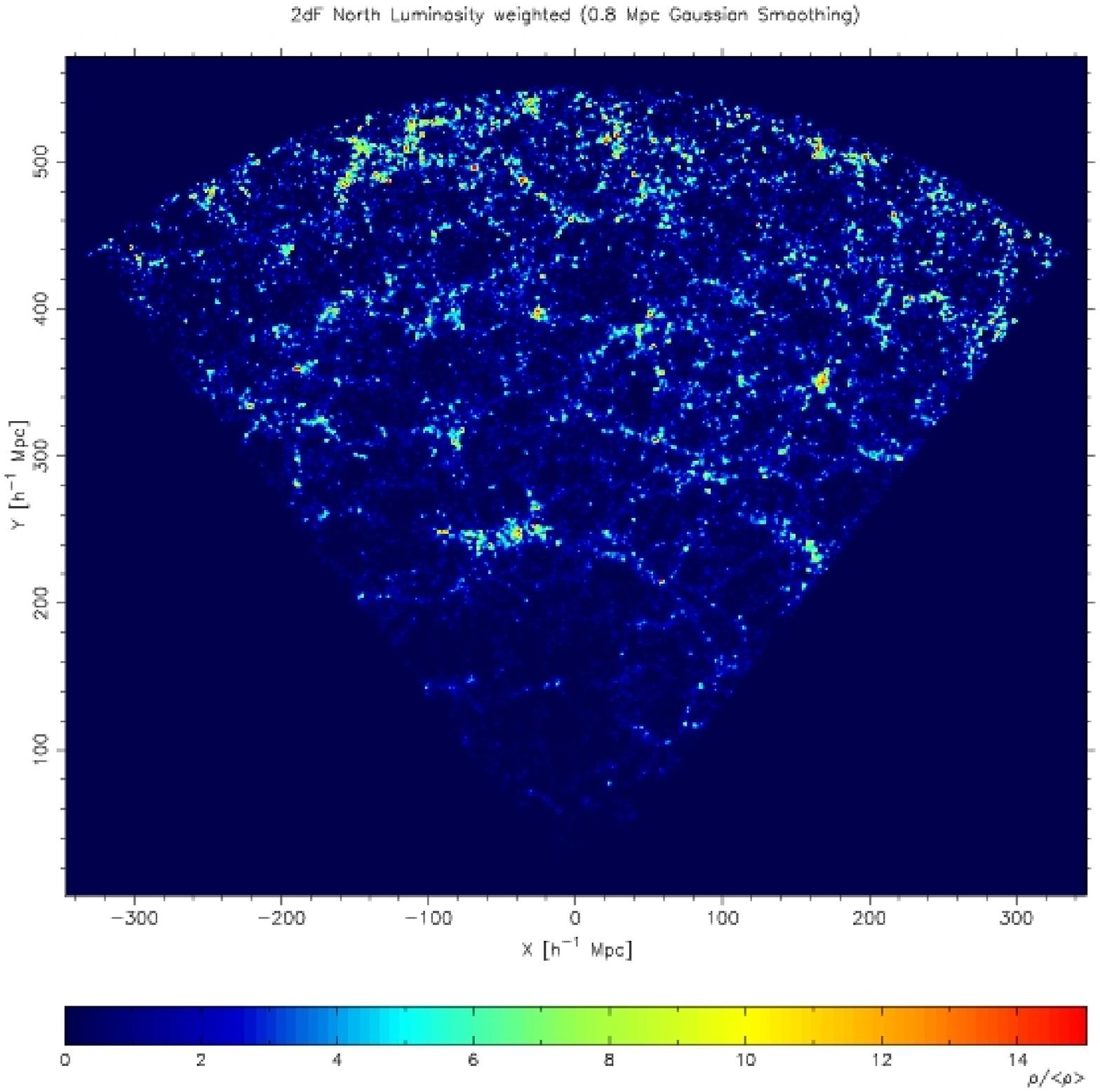}}
\resizebox{0.49\textwidth}{!}{\includegraphics*{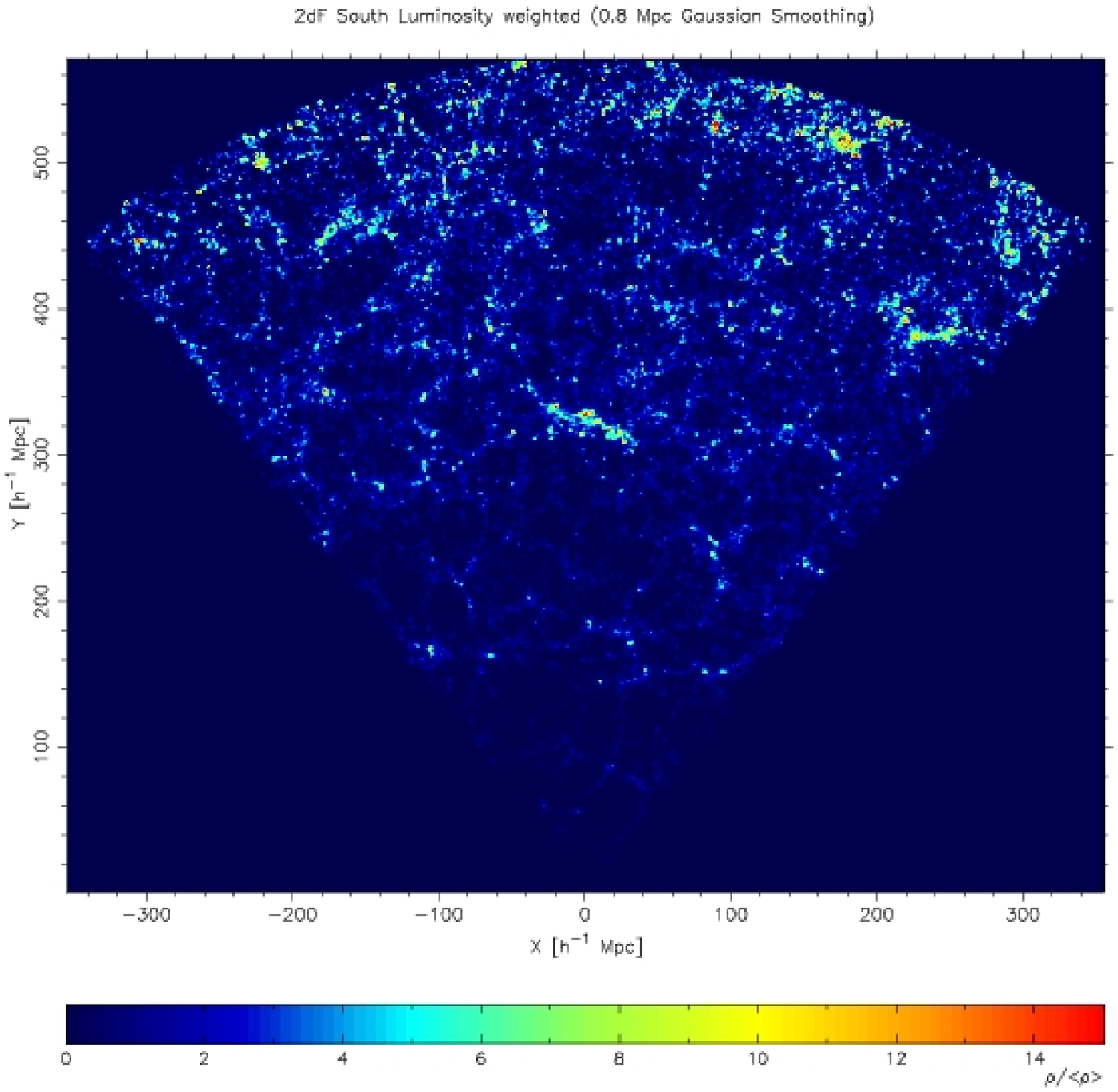}}\\
\caption{The high-resolution 2-dimensional density fields of the
  Northern and Southern parts of the 2dF redshift survey are shown in the left
  and right panels, respectively.  The samples are conical, i.e. their
  thickness increases with distance, thus at large distances from the observer
  we see many more systems of galaxies. The richest supercluster in the
  Northern region is SCL126 in the list by Einasto et al. \cite{e1997}, also
  called the Sloan Great Wall by Vogeley et al. \cite{vogeley04}; the richest
  Southern supercluster is SCL9 by Einasto et al. \cite{e1997}, or the
  Sculptor Supercluster.}
\label{fig:1}
\end{figure*}

\section{Superclusters in the 2dF and Sloan surveys and in the Millennium
  simulation} 

Both real and model superclusters were found using the luminosity density
fields calculated using Epanechnikov smoothing with a radius of 8~\Mpc.
Superclusters were defined as connected non-percolating systems with densities
above a certain threshold density.  These density fields were normalized to
identical mean levels, and all regions above a threshold density 6 (in units
of the mean density) were considered as superclusters.  The density fields
were calculated for a grid step of 1~\Mpc, which allows to investigate the
detailed spatial structure of superclusters.  The 2dFGRS superclusters were
found separately for the Northern and Southern regions of the 2dF Survey, and
SDSS superclusters -- for the high-declination region of the SDSS DR4 Survey in
the Northern hemisphere.  The 2dF Northern and Southern regions together
contain 544 superclusters, the SDSS Northern survey has 911 superclusters. The
comparison model samples have 1733 and 1068 superclusters (the full model
sample and the simulated 2dF sample, respectively).

\begin{figure*}[ht]
\centering
\resizebox{0.48\textwidth}{!}{\includegraphics*{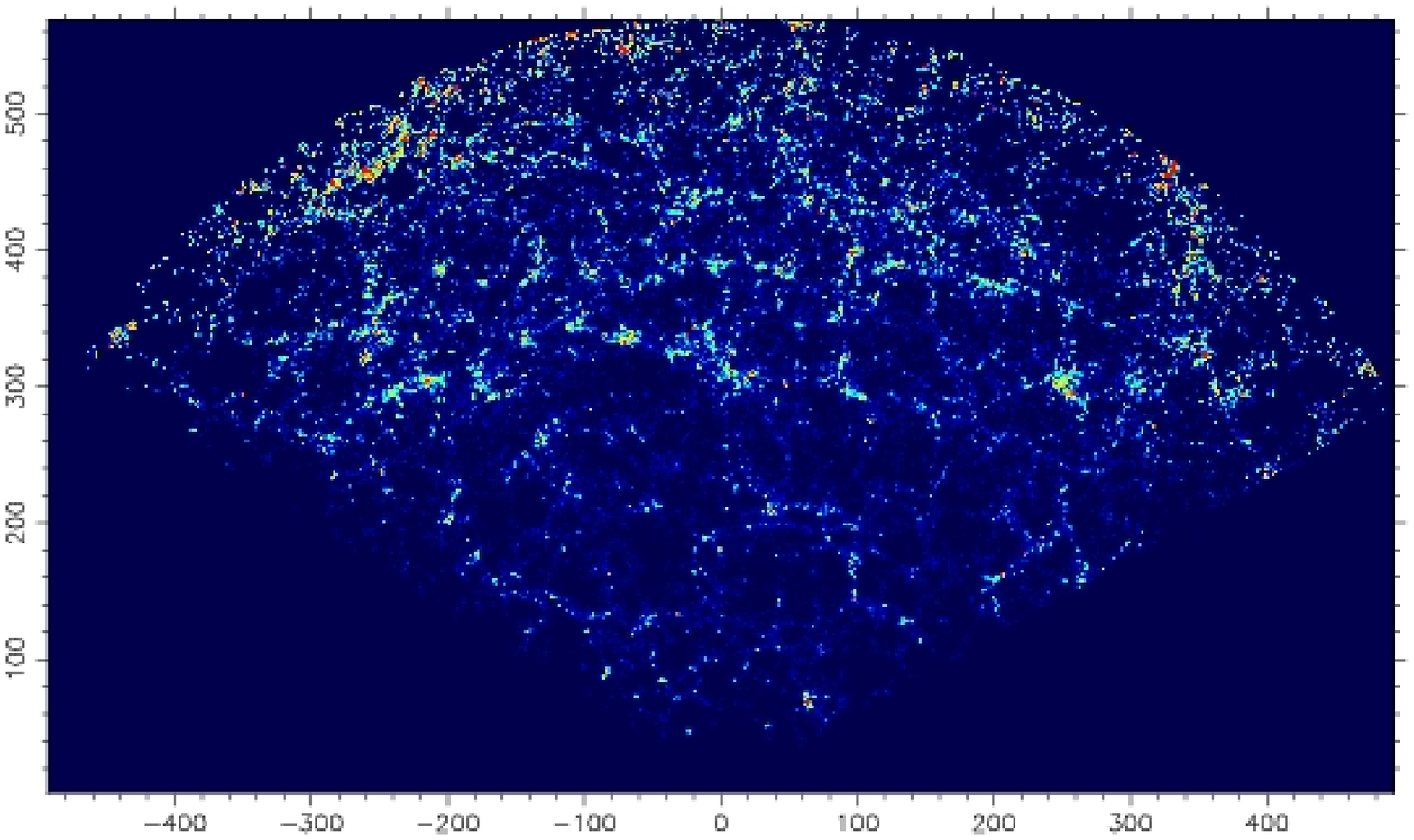}}\\
\resizebox{0.48\textwidth}{!}{\includegraphics*{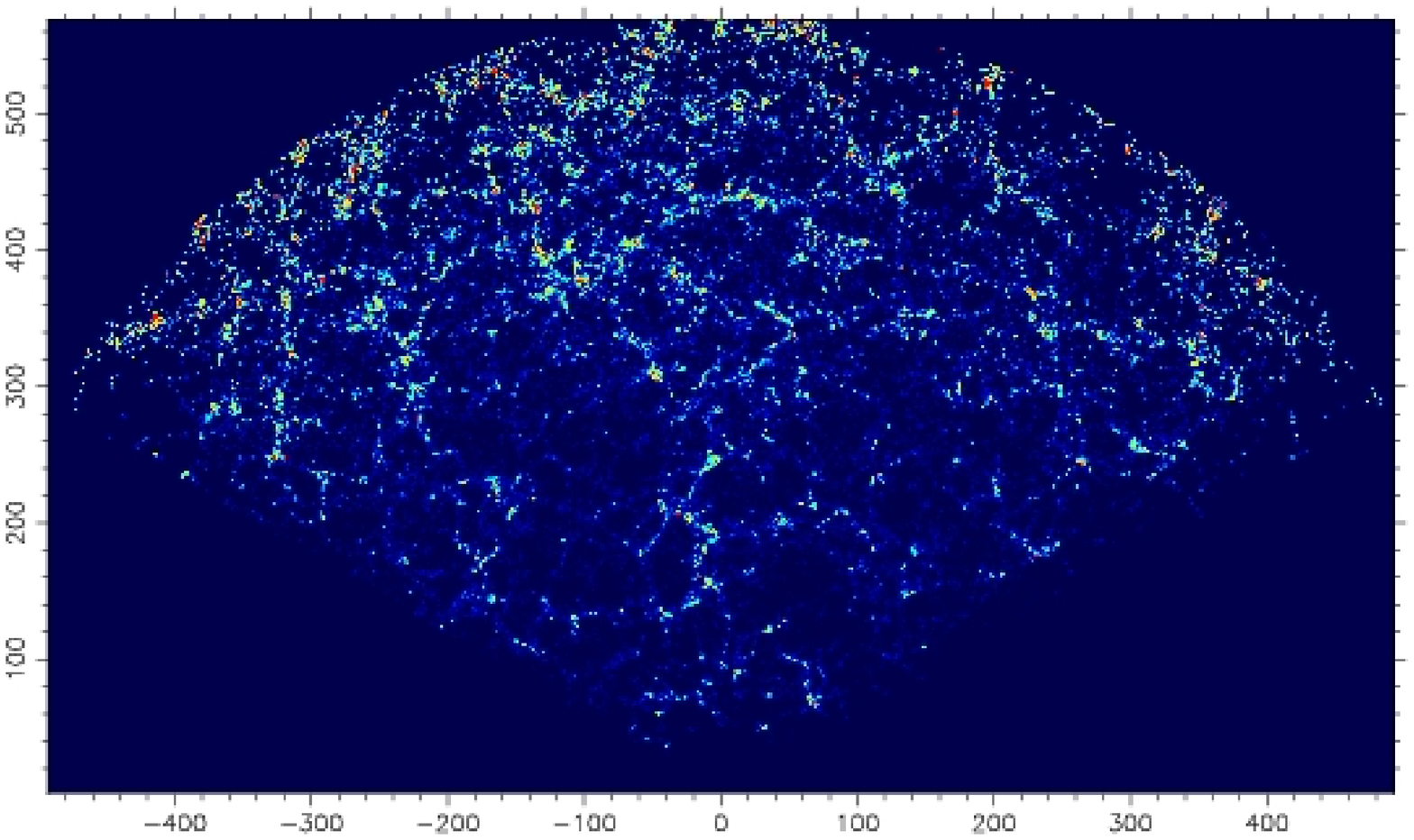}}\\
\resizebox{0.48\textwidth}{!}{\includegraphics*{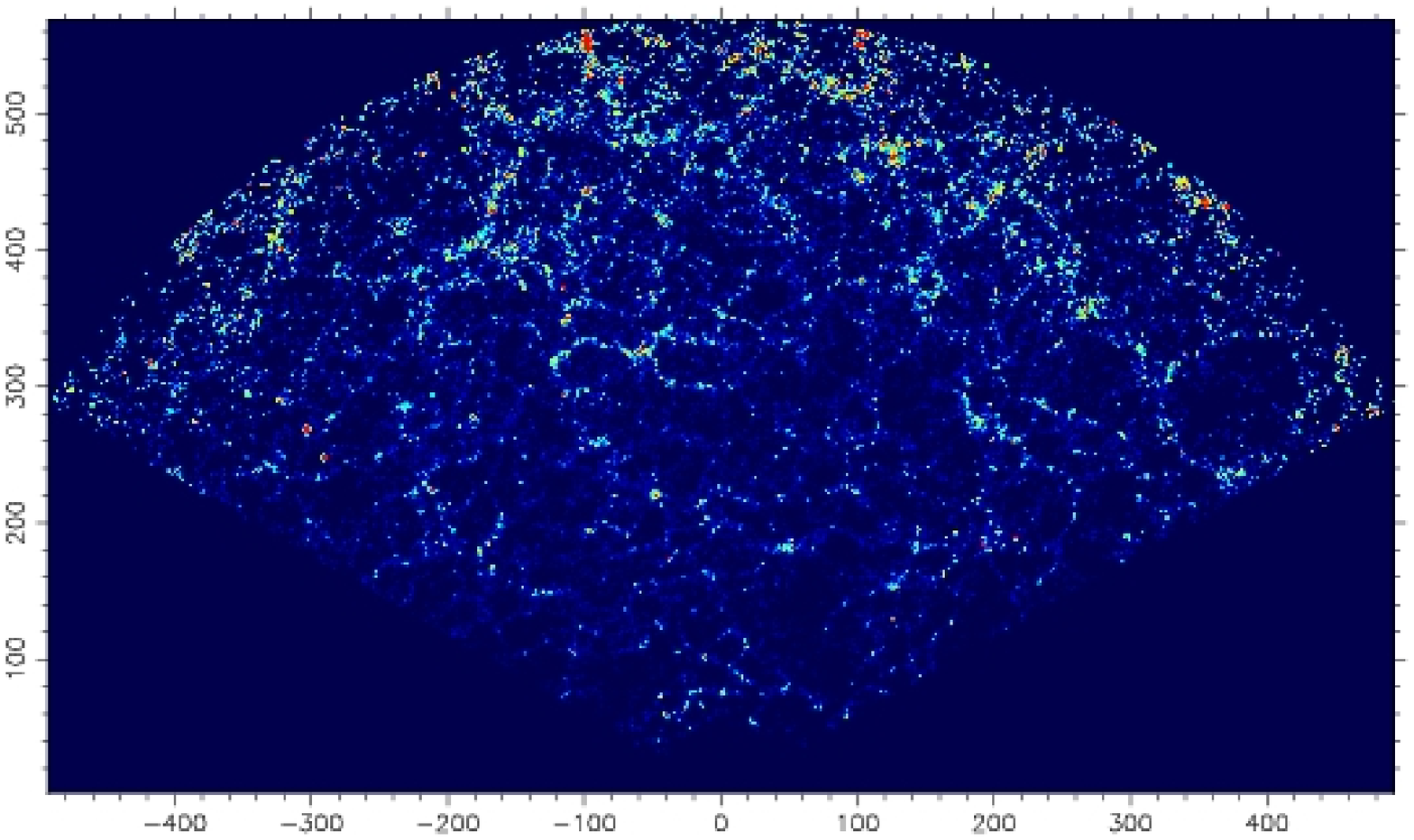}}
\caption{The high-resolution 2-dimensional density field of the
  SDSS DR4 Northern high-declination region. The wedges are drawn in
  rectangular coordinates based on the SDSS survey coordinates $\eta$ and
  $\lambda$, and have a thickness of 9.33 degrees in $\lambda$; the panels
  from bottom to top correspond to increasing $\lambda$ values. Note the
  presence of rich superclusters in all wedges.  }
\label{fig:2}
\end{figure*}

In Figs.~\ref{fig:1} and \ref{fig:2} we show high-resolution density fields on
the 2dFGRS and SDSS surveys. All  wedges are about 10 degrees
thick, thus near to the observer they are thin.  These Figures show the cosmic
web -- a continuous network of galaxy systems of various luminosity densities,
and voids between them.  All luminous regions seen in these Figures are
superclusters.  We see that they have very different richness, some are very
small and resemble the Local Supercluster around the Virgo cluster, some are
large and very rich.

For all superclusters their geometric and physical properties were found.
Among the geometrical properties are the position (RA, DEC and distance), the
size, and the offset of the geometrical center from the dynamical one, defined
as the center of the main (most luminous) cluster.  The physical properties
are the mean and maximum luminosity densities, the total luminosity and the
luminosity of the main cluster and of the main galaxy (the brightest galaxy of
the main cluster).

Comparison of properties of model superclusters with properties of
real superclusters shows that they are very similar.  Superclusters consist of
several chains (filaments) of galaxies, groups and clusters. These chains have
various length, thus superclusters are asymmetrical in shape. The degree of
asymmetry is higher in rich superclusters.  Rich superclusters are also denser
and contain luminous knots -- high-density nuclei.

\section{Rich superclusters in real data and models}

One important property of superclusters is different in real and model samples
-- the supercluster richness.  To characterise the richness we used two
independent characteristics: the total luminosity and the number of rich
clusters, i.e. the multiplicity.

The multiplicity was derived from the number of high-density knots of the
density field. We call these knots DF-clusters.  The spatial density of
DF-clusters is about two times higher than the spatial density of Abell
clusters in the same volume, thus the expected number of DF-clusters in
superclusters is about two times higher than the number of Abell clusters.
Both functions were determined separately for the 2dFGRS and SDSS
superclusters, and for the total observational sample.

\begin{figure*}[ht]
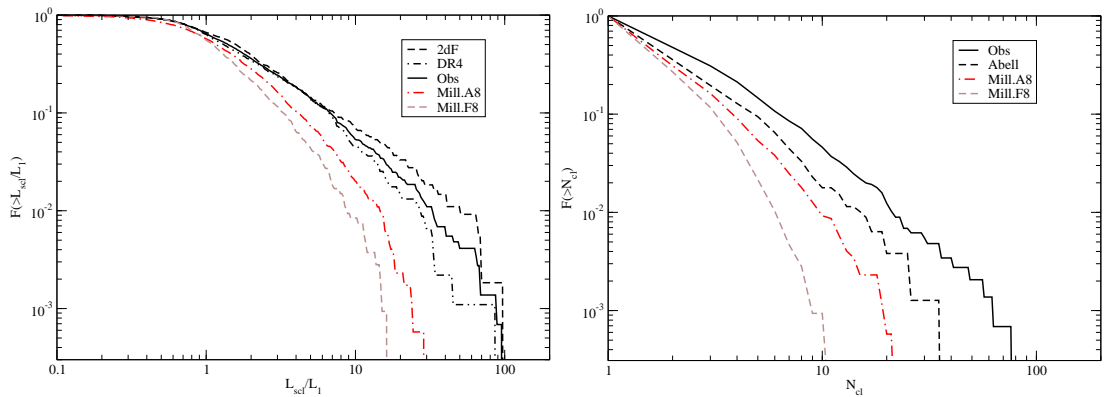

\centering
\resizebox{0.48\textwidth}{!}{\includegraphics*{scl6.0_Obs-Mill_lumf.eps}}
\resizebox{0.48\textwidth}{!}{\includegraphics*{scl6.0_Obs-Mill_multf.eps}}
\caption{
  Comparison of relative luminosity functions and multiplicity functions of
  observational and model supercluster samples (in left and right panels,
  respectively). The spatial density of
superclusters is expressed in terms of the total number of
superclusters in a sample to avoid small differences due
to different mean supercluster number densities in different samples. In
the left panel we show the relative luminosity functions separately for
the observational samples SDSS DR4, 2dF, and for the combined sample Obs, in
the right panel we use the combined observational sample Obs, and the Abell
supercluster sample (here the multiplicity is defined by the number of Abell
clusters, and isolated Abell clusters are considered as richness class 1
superclusters). Mill.A8 and Mill.F8 denote the full and simulated 2dF
  Millennium supercluster samples, respectively.}
\label{fig:3}
\end{figure*}

The total luminosity was calculated by summing the luminosities of galaxies
and clusters of galaxies inside the density contour, which defines the
boundary of the supercluster. In our case the threshold density was chosen to
be 6 (in units of the mean density).  In calculating total luminosities we
used weights for galaxies which take into account the galaxies outside the
observational luminosity window of a survey.  To avoid complications due to
the use of different color systems and mean luminosities, we used {\em
  relative} luminosities, normalized by the mean luminosity of poor
superclusters, i.e. the superclusters that contain only one DF-cluster.

For model samples we calculated these functions for two cases.  One sample
uses all model galaxies and can be considered as the ''true'' model sample.
The second model sample simulates the 2dF sample, where an ''observer'' was
put into one corner of the sample, and only these galaxies were included,
which satisfy the same selection criteria as used in the real 2dF sample.

The luminosity and multiplicity functions of real and model samples are
compared in Fig.~\ref{fig:3}.  We see that both functions show for real
samples much more rich superclusters than for model samples.  This difference
is the major cosmological result of the analysis of our supercluster survey.
The presence of very rich superclusters in our vicinity is well known; good
examples are the Shapley Supercluster and the Horologium-Reticulum
Supercluster (see Fleenor et al.  \cite{fleenor05}, Proust et al.
\cite{proust06}, Nichol et al.  \cite{nichol06} and Ragone et al.
\cite{ragone06} and references therein).  But until recently the number of
such extremely massive superclusters was too small to make definite
conclusions about their abundance.

When comparing models with observations we have to use the simulated 2dF
sample, which is formed using the same selection criteria as used for the
observational sample.  The most luminous simulated superclusters of this
sample have a relative luminosity of about 15 in terms of the mean luminosity
of richness class 1 superclusters, whereas the most luminous superclusters of
real samples have a relative luminosity about 100, i.e. they are about 6 times
more luminous.  The richest model superclusters have a multiplicity of 10,
whereas the multiplicity of the richest real superclusters is over 70.  The
number of Abell clusters in the richest Abell supercluster is 34 (Einasto et
al. \cite{e2001}).

Figures \ref{fig:1} and \ref{fig:2} show that very luminous superclusters are
located in {\em all subsamples} (the Northern and Southern regions of the
2dFGRS, and in subregions of the SDSS DR4 sample, if divided into 3 wedges of
equal width).  These subsamples have characteristic volumes of about 10
million cubic \Mpc, whereas model samples of 10 times larger volume have no
extremely rich superclusters.

\section{Formation of the supercluster-void network}

To check these results we used a number of independent numerical
simulations, carried out for simulation boxes of size of 500~\Mpc\ and
768~\Mpc, using $512^3$ Dark Matter particles.  We  found DM-halos
in simulations, and used them to calculate the smoothed density field as
for real and Millennium Simulation.  For all simulations we then found
simulated superclusters as previously, and found the distribution of
dense knots (simulated rich clusters).  These calculations confirmed
our previous result: the number of dense knots in simulated
superclusters is much lower than in real superclusters.

This striking conflict between model and reality needs explanation. In order
to understand the formation of rich superclusters we used wavelet analysis to
investigate the role of density waves of different scales.  The \'a trous
wavelet technique we used allows to divide the density field into components
of various wavelength bands, so that the field is restored by summing all
components.  The wavelet analysis was carried out both for real and model
samples.

Our results show that in all cases superclusters form only in
regions where {\em large density waves combine in similar local phases to
generate high density peaks}.  Very rich superclusters are objects
where density waves of all large scales (up to a wavelength $\sim 250$~\Mpc)
have similar phases.  The smaller is the maximum wavelength of such
phase synchronization, the lower is the richness of superclusters.
Similarly, large voids are caused by large-scale density perturbations
of wavelength $\sim 100$~\Mpc, here large-wavelength modes combine
{\em in similar local phases to generate under-densities}.

Superclusters of galaxies are formed by density perturbations of large scales.
These perturbations evolve very slowly.  As shown by Kofman \& Shandarin
\cite{kofman88}, the present structure on large scales is built-in already
in the initial field of linear gravitational potential fluctuations.  Actually
they are remnants of the very early evolution and stem from the inflationary
stage of the Universe (see Kofman et al. \cite {kofman87}.  The distribution
of luminosities of superclusters allows us to probe processes
acting at these very early phases of the evolution of the Universe.

There are two possible explanations for the large difference between the
distribution of luminosities of real and simulated samples.  One possibility
is that in present simulations the role of very large density perturbations,
responsible for the formation of these very luminous superclusters, is
underestimated.  The other feasible  explanation of the differences between
models and reality may be the presence of some unknown processes in the
very early Universe which give rise to the formation of extremely luminous
and massive superclusters.

\section{Conclusions}

\begin{enumerate}

\item Geometric properties of superclusters are well explained by current
  models.

\item There are much more very rich superclusters than models predict.

\item Large perturbations evolve very slowly and represent the fluctuation
  field at the epoch of inflation.

\item The difference between observations and models can be explained in two
  ways: \\
  large-scale perturbations are not incorporated in the models, i.e. models
  need   improvement; \\
  there ocurred presently unknown processes during  inflation.

\end{enumerate}

The present review is based on talks held in Budapest on April 20,
2006 in Detre Centenarium, in Uppsala University on April 27, 2006 and
in Aspen Workshop on Cosmic Voids on June 6, 2006.  I thank my
collaborators Maret Einasto, Enn Saar, Erik Tago and Volker M\"uller
for permission to use results of our common work in this review.  We
are pleased to thank the 2dFGRS and SDSS Teams for the publicly
available data releases.  The present study was supported by Estonian
Science Foundation grants No.  4695, 5347 and 6104, and Estonian
Ministry for Education and Science support by grant TO 0060058S98.  I
thank Astrophysikalisches Institut Potsdam (using DFG-grant 436 EST
17/2/05) and Uppsala University for hospitality where part of this
study was performed.  2dFGRS supercluster catalogues are available at
{\tt http://www.aai.ee/~maret/2dfscl.html}, Sloan DR4 supercluster
catalogues at {\tt http://www.aai.ee/~maret/SDSSDR4scl.html}.


\begin{thebibliography}{}


\bibitem[(1958)]{abell} Abell, G., 1958, ApJS, 3, 211

\bibitem[(1961)]{abell61}  Abell, G., 1961, AJ, 66, 607

\bibitem[(1989)]{aco} Abell, G., Corwin, H., Olowin, R., 1989, ApJS, 70, 1

\bibitem[(2006)]{dr4} Adelman-McCarthy, J.K., Ag\"ueros, M.A., Allam, S.S. et
  al. 2006, ApJS, 162, 38

\bibitem[(2003)]{bas03} Basilakos, S., 2003, MNRAS, 344, 602


\bibitem[(2001)]{bpr01} Basilakos, S., Plionis, M., Rowan-Robinson, M.,
  2001, MNRAS, 323, 47

\bibitem[(2001)]{col01} Colless, M.M., Dalton, G.B., Maddox, S.J., \etal,
     2001, MNRAS, 328,  1039

\bibitem[(2003)]{col03} Colless, M.M., Peterson, B.A., Jackson, C.A., \etal,
     2003, (astro-ph/0306581)

\bibitem[(2006)]{croton06} Croton, D.J., Springel, V., White, S.D.M. et
  al. 2006, MNRAS, 365, 11

\bibitem[(1953)]{deV53} de Vaucouleurs, G., 1953, AJ, 58, 30

\bibitem[(2003a)]{e03a} Einasto, J., Einasto, M.,  H\"utsi, G., \etal,~
     2003a,  A\&A, 410, 425 (E03a)

\bibitem[(2006b)]{e06b} Einasto, J., Einasto, M., Saar, E. et al. 2006b, A\&A,
     (accepted, Paper II, astro-ph/0604539)

\bibitem[(2006c)]{e06c} Einasto, J., Einasto, M., Saar, E. et al. 2006b, A\&A,
     (accepted, astro-ph/0605393)




\bibitem[(2006a)]{e06a} Einasto, J., Einasto, M., Tago, E. et al. 2006a, A\&A,
  (submitted, Paper I, astro-ph/0603764)


\bibitem[(2003b)]{e03b} Einasto, J.,  H\"utsi, G., Einasto, M., \etal,~
     2003b,  A\&A, 405, 425

\bibitem[(2005)]{ein05} Einasto, J., Tago E., Einasto, M., et al. 2005,
   A\&A, 439, 45


\bibitem[(2001)]{e2001} Einasto, M., Einasto, J., Tago, E., M\"uller, V.
\& Andernach, H.,  2001, AJ, 122, 2222

\bibitem[(1997)]{e1997} Einasto, M., Tago, E., Jaaniste, J., Einasto, J.
\& Andernach, H., 1997, A\&A Suppl., 123, 119 (E97)

\bibitem[(2004)]{erd04} Erdogdu, P., Lahav, O., Zaroubi, S. et al. 2004,
  MNRAS, 352, 939

\bibitem[(2005)]{fleenor05} Fleenor, M.C., Rose, J.A., Christiansen,
  W.A. et al. 2005, AJ, 130, 957



\bibitem[(1978)]{gt78} Gregory, S.A. \& Thompson, L.A. 1978, ApJ, 222, 784

\bibitem[(1978)]{jet78} J\~oeveer, M., Einasto, J. \& Tago, E. 1978,
  MNRAS, 185, 357


\bibitem[(1987)]{kofman87} Kofman, L.A., Linde, A.D. \& Einasto, J. 1987,
  Nature, 326, 48

\bibitem[(1988)]{kofman88} Kofman, L.A. \& Shandarin, S.F. 1988, Nature, 334,
  129


\bibitem[(2006)]{nichol06} Nichol, R.C., Sheth, R.K., Suto, Y., et
  al. 2006, MNRAS, 368, astro-ph/0602548


\bibitem[(2005)]{pr05} Porter, S.C. \& Raychaudhury, S. 2005, MNRAS,
  364, 1387


\bibitem[(2006)]{proust06} Proust, D., Quintana, H., Carrasco, E.R. et
  al. 2006, A\&A, 447, 133

\bibitem[(2006)]{ragone06} Ragone, C.J.,  Muriel, H., Proust, D. et al.
2006, A\&A, 445, 819



\bibitem[(2005)]{springel05} Springel, V., White, S.D.M., Jenkins, A. et
  al. 2005, Nature, 435, 629

\bibitem[(2006)]{tago06} Tago, E., Einasto, J.,  Saar, E. et al. 2006,
  AN, 327, 365  (T06)

\bibitem[(2004)]{vogeley04} Vogeley, M.S., Hoyle, F., Rojas, R.R. et
  al. 2004, Proc. IAU Coll. No. 195 (astro-ph/0408583)



\bibitem[(1982)]{zes82}  Zeldovich, Ya.B., Einasto, J.,
Shandarin, S.F., 1982, Nature, 300, 407







\end{thebibliography}
\end{document}